# A simple patchy colloid model for the phase behavior of lysozyme dispersions

Christoph Gögelein,[1,a)] Gerhard Nägele,[1] Remco Tuinier,[1] Thomas Gibaud,[2] Anna Stradner,[2] and Peter Schurtenberger[2]
[1]*Institut für Festkörperforschung, Teilinstitut Weiche Materie, Forschungszentrum Jülich, D-52425 Jülich, Germany*
[2]*Physics Department and Fribourg Center for Nanomaterials, University of Fribourg, CH-1700 Fribourg, Switzerland*

We propose a minimal model for spherical proteins with aeolotopic pair interactions to describe the equilibrium phase behavior of lysozyme. The repulsive screened Coulomb interactions between the particles are taken into account assuming that the net charges are smeared out homogeneously over the spherical protein surfaces. We incorporate attractive surface patches, with the interactions between patches on different spheres modeled by an attractive Yukawa potential. The parameters entering the attractive Yukawa potential part are determined using information on the experimentally accessed gas-liquid-like critical point. The Helmholtz free energy of the fluid and solid phases is calculated using second-order thermodynamic perturbation theory. Our predictions for the solubility curve are in fair agreement with experimental data. In addition, we present new experimental data for the gas-liquid coexistence curves at various salt concentrations and compare these with our model calculations. In agreement with earlier findings, we observe that the strength and the range of the attractive potential part only weakly depend on the salt content.

## I. INTRODUCTION

The exploration of crystallization processes of proteins has been the subject of active research since obtaining regular crystals is indispensable for the structural analysis using, e.g., x-ray scattering tools.[1] In practice, crystallographers need to screen many batches by varying the solution properties until the proper conditions are found where regular crystals are formed.[2] Obviously, this approach is time consuming and tedious, and one would like to have a rule of thumb to know in advance as to what conditions a successful crystallization route may be achieved. Several ways to accelerate structural analysis have been discussed, e.g., transferring the proteins to solvent conditions far away from their native environment by increasing the salt concentration (salting-out effect), adding di- and multivalent ions (Hofmeister series), and varying the $p$H-value, the temperature, or adding depletion agents.[1]

The application of concepts from colloidal science to proteins has led to progress in understanding their interactions and phase behavior. By applying the Derjaguin–Landau–Verwey–Overbeek (DLVO) theory of colloidal stability[3] to proteins, and by adjusting the van der Waals interaction to match the experimental data, it had been concluded that proteins interact essentially by long-ranged screened electrostatic repulsion due to their effective surface charges, and by short-ranged attractive forces responsible for a metastable gas-liquid coexistence curve.[4–6] In addition, the adhesive hard-sphere model, as exemplified by the sticky-sphere model, has been applied to protein solutions.[7–9] However, in the presence of such extremely deep ($\sim 8 k_B T$) and short-ranged attractions ($\sim 10\%$ of the protein diameter) obtained from models with isotropic interactions using the DLVO theory, one might expect that the proteins coagulate, whereas noncoagulated stable phases are observed.[10]

Experiments on the phase behavior have been focused so far mainly on solutions of lysozyme proteins. For these systems, a large amount of data and insight has been accumulated during the past: Taratuta et al.[11] have discovered the existence of a gas-liquid coexistence curve, which was subsequently shown to be metastable with respect to the fluid-crystal phase separation by Broide et al.[10] George and Wilson[12] have found that there is a narrow band of negative values for the second virial coefficient for which crystallization occurs. Thereafter, ten Wolde and Frenkel[13] demonstrated that the nucleation barrier is lowered in the region close to the critical point. As a consequence, the understanding and prediction of the fluid phase behavior has turned out to be a prerequisite to describe nucleation kinetics. For a more detailed general discussion of protein crystallization, we refer to the two reviews by Piazza in Refs. 8 and 14.

Further progress in explaining the experimental gas-liquid phase separation was made by considering anisotropic protein interactions. To investigate the influence of attractive patches on the protein surfaces, Lomakin et al.[15,16] have used an orientation-dependent square-well potential, which allows for a remarkably good description of the gas-liquid phase coexistence as well as for the solubility curve. Moreover, they demonstrated that whether one is allowed to orientationally average the angular-dependent pair potential depends

[a)]Electronic mail: c.goegelein@fz-juelich.de.

strongly on the number of nearest neighbors and on the number and size of patches. Thus, taking into account the anisotropic interactions is crucial in describing crystallization in lysozyme solutions.

Kern and Frenkel[17] have discussed the phase behavior by accounting for the relative orientation of two interacting molecules. Different from them, Lomakin et al.[15,16] disregarded in their computer simulation study the anisotropy of all surrounding particles in the total interaction pair potential. From their computer simulations, Kern and Frenkel conclude that the critical temperature decreases as the surface area of the attractive patches decreases. Moreover, from their simulations follows that the critical volume fraction depends only weakly on the patch area, and that at constant surface coverage, the critical temperature decreases with decreasing number of patches. According to Kern and Frenkel, the critical point is no longer characterized by a unique value of the second virial coefficient but rather depends on the number and area of patches.

Recently, Liu et al.[18] have extended the approach of Kern and Frenkel on assuming a sum of a patchy and an isotropic square-well attraction. They find good agreement between the experimental gas-liquid coexistence curve and their theoretical binodal. In their model, a heuristic set of interaction parameters determining the range and strength of the isotropic and anisotropic interaction potential part is chosen by scaling the temperature and the particle density with the experimental values at the critical point. Additionally, they observe that the location of the critical point is only slightly affected by the surface distribution of patches.

Sear[19] has approached the problem of protein crystallization by applying Wertheim's perturbation theory[20,21] for self-associating fluids, and he obtains a qualitative description of the phase behavior. In Wertheim's theory, the interactions are assumed pointlike, so only site-site bounds can be formed. Clusters and percolated gels are described here by assuming nonvanishing probabilities for the formation of monomers, dimers, and so on, leading to a statistical description of the associating fluid. This approach has afterward been used by Warren[22] to explore the influence of the added salt on the phase behavior of lysozyme. In addition, Sear's model has been used by Zukoski and co-workers, to address the problem of the nucleation kinetics in protein solutions (see Ref. 23 and references therein). They have also compared their results to the experimental data on crystal nucleation kinetics.[23,24]

Despite this success and the valuable insight gained by using Wertheim's perturbation theory, the Sear model lacks the incorporation of patches. Fantoni et al.[25] pointed to this shortcoming of the Sear model, and developed an analytical description for patchy hard spheres using Baxter's adhesive sphere model. They compared their results for the structure in the anisotropic liquid and the equilibrium phase behavior with their computer simulations.

An anisotropic interaction-site lattice model was proposed by Talanquer.[26] In this work, the occurrence of nonspherical critical nuclei is predicted, whose specific geometry depends on the strength of the anisotropic interactions.

The influence of the number of patches on the crystal lattice structure has been investigated by Chang et al.[27] using computer simulation methods. Interestingly, in the case of a model with six patches, they observe a phase transition from a simple cubic (sc) to an orientationally disordered face-centered-cubic lattice (fcc) above room temperature. In addition, they observed a metastable transition between the orientationally disordered and ordered fcc lattice at lower temperature. This study demonstrates that anisotropic interactions can lead to manifold crystal structures depending crucially on the geometry and strength of the patchy interactions.

Quite recently, theoretical work on dispersions of patchy colloid particles has caused much attraction due to the progress made by Bianchi et al.[28] On varying the patchiness, they demonstrated that patchy colloids can offer the possibility to generate a beforehand inaccessible liquid state, with a possible percolation threshold at temperatures below the critical point without a preceding gas-liquid phase separation.

Common to all previous studies incorporating anisotropic interactions is that they use a square-well potential to describe the attractive interaction part between the proteins. A square-well form, however, is only realistic in case of a very short-ranged attraction and negligible nonexcluded volume repulsions such as in high-salt systems. On decreasing the salinity, the range of the screened electrostatic repulsion increases. Hence, the fluid phase becomes stabilized against gas-liquid phase separation on lowering the salt content, and one can expect that the critical point is shifted to lower temperatures. For zero added salt, one expects in lieu of a gas-liquid coexistence a microphase separation to take place,[29–31] which actually has been seen experimentally.[32] Such equilibrium clusters form if, first, the range of repulsion is large enough to stabilize the conglomerates against further growth, and second, if the attractive forces are sufficiently short-ranged to hinder particles from escaping the cluster.

To investigate the influence of discrete charge patterns on the protein surfaces regarding many-body interactions, Allahyarov et al.[33] have performed molecular dynamics simulations where, in addition, the finite size of the microions has been accounted for. They observe deviations in the angular-averaged pair potential from the monotonic decaying behavior predicted by DLVO theory for large ionic strengths in lysozyme solutions.

The thermodynamic properties of lysozyme crystals have been investigated in detail by Chang et al.[34] They have combined atomistic Monte Carlo simulation to account for the anisotropic shape and van der Waals attractions with a boundary element method solving the Poisson–Boltzmann equation to account for the discrete charge distribution close to the lysozyme surface and the effect of salt-induced screening. Whereas the predicted van der Waals energy and the electrostatic energy are in good agreement with experimental data for a tetragonal lattice structure, poorer agreement is found for an orthorhombic lattice structure. This discrepancy can be attributed to both a change in the solvation structure, which has been observed experimentally, and to the general difficulties in describing van der Waals interactions quantitatively.

In this paper, we include the screened electrostatic repulsion explicitly to separate the influence of these Coulomb interactions from the attractive forces in lysozyme solutions, which are presumably induced by hydrophobic interactions and dispersion forces.[35,36] In our model calculations, the patchy attractive forces are assumed to be of a Yukawa-type form. This enables us to characterize and to quantify the strength and range of the radial attractive pair potential part from the experimental critical point and the measured binodals, as well as to investigate the competitive effect of repulsive and attractive pair forces on the phase behavior as a function of salinity.

This paper is organized as follows. In Sec. II, we describe the sample preparation and the experimental techniques used to obtain the phase diagram. To model the attractive patchy pair interactions, we factorize its angular and radial degrees of freedom using the patchy model of Kern and Frenkel,[17] assuming an attractive Yukawa potential for the radial factor (see Sec. III). The Helmholtz free energy of the fluid and solid phases is calculated using the second-order perturbation theory, as described in Sec. IV. In Sec. V, we explain how we determine the range and strength of the attractive potential part, as well as the patchiness of the proteins, using information on the experimentally observed critical point. For this purpose, we use an earlier finding of Warren[22] on the second osmotic virial coefficient of lysozyme solutions, and an extended corresponding state argument of Noro and Frenkel.[37] This simplifying strategy enables us to quantify the range and strength of attraction, and the surface area fraction covered by attractive patches. In Sec. VI, we present the calculated phase diagrams. To compare the theoretical coexistence curves with the experimental data, we include the temperature dependence of the attractions. In Sec. VII, we discuss the so obtained physical parameters in comparison to previous findings. The capability of our model to describe the influence of the added salt on the gas-liquid coexistence curve is demonstrated through a comparison with the existing[38] and new experimental data on lysozyme solutions at various salinities. We also predict the fluid-solid coexistence curve for the experimentally given salt concentrations. Finally, in Sec. VIII, we present our conclusions.

## II. EXPERIMENTAL DETAILS

We have used hen egg-white lysozyme purchased from Fluka, Inc. (L7651). The molar mass of lysozyme is 14 400 g/mol and its mass density is $\rho_0 = 1.351$ g/cm$^3$. In experiments, proteins have been dissolved with a $c_b = 0.02$ mol/l (2-hydroxyethyl)piperazine-$N'$-(2-ethanesulfonic acid) (HEPES) buffer solution without the added salt. The $p$H has been adjusted to $7.8 \pm 0.1$ using a sodium hydroxyl solution.[39,40] At this $p$H-value, it is known from titration experiments that the protein carries $Z=8$ net positive elementary charges.[41] The stock solution has been diluted with a buffer solution containing sodium chloride to the desired volume fraction and excess salt concentration. Partial phase separation was avoided by mixing the buffer and stock solution at temperatures well above the gas-liquid coexistence curve. In this way, transparent samples at room temperature have been prepared with a protein volume fraction in the range from 0.01 to 0.18. The concentrations have been measured by ultraviolet absorption spectroscopy using a specific absorption coefficient ($E_{1\ \text{cm}}^{1\%} = 26.4$). Highly concentrated samples at volume fractions up to 34% have been prepared by quenching a solution, typically of 15.5% volume fraction, to temperatures in the range 15 °C $<T<$ 18 °C below the cloud point, and centrifuging the system for 10 min at 9 $\times 10^3$ g. The highly concentrated bottom phase has been used for further experiments.

We have determined the binodal curve by cloud point measurements. A sample of given volume fraction was placed in a temperature-controlled water bath well above the critical point. Then, the temperature was slowly decreased, and the cloud point determined by the temperature where the solution turns turbid. The spinodal temperature and the critical point have been estimated by static light scattering measurements at 90° scattering angle using a 3D light scattering setup (LS-Instruments GmbH, $\lambda=633.6$ nm).[42]

## III. MODEL

We assume that the total pair potential, $u(r, \Omega_1, \Omega_2)$, between two spherical proteins at a center-to-center distance $r$ can be described by a known repulsive isotropic interaction potential part $u_{\text{rep}}(r)$ due to the effective charges on the protein surfaces and an attractive, patchy interaction part $u_{\text{attr}}(r, \Omega_1, \Omega_2)$ with yet unspecified interaction parameters. The finite size of the spherical protein is accounted for using a hard-sphere potential $u_0(r)$ by mapping the ellipsoidal-like shape[43] of a lysozyme protein onto an effective sphere, as explained at the end of this section. In total,

$$u(r, \Omega_1, \Omega_2) = u_0(r) + u_{\text{rep}}(r) + u_{\text{attr}}(r, \Omega_1, \Omega_2). \quad (1)$$

Here, $\Omega_i$ is the solid angle of a sphere $i$, and the hard-sphere potential part is

$$u_0(r) = \begin{cases} \infty, & r \leq \sigma \\ 0, & r > \sigma, \end{cases} \quad (2)$$

where $\sigma$ denotes the protein diameter.

The repulsive pair interaction part is described by the electrostatic part of the one-component macroion-fluid potential,[44]

$$\beta u_{\text{rep}}(r) = \begin{cases} Z^2 l_B Y^2 \dfrac{\exp[-z_{\text{rep}}(r/\sigma - 1)]}{r}, & r > \sigma \\ 0, & r \leq \sigma, \end{cases} \quad (3)$$

Here, $Z$ is the protein charge number and $l_B = e^2/(4\pi\varepsilon_0\varepsilon k_B T)$ is the Bjerrum length with the dielectric constant *in vacuo* $\varepsilon_0$, the dielectric solvent constant $\varepsilon$, and the elementary charge $e$.

The effect of the finite size and concentration of the colloidal macroions is incorporated by the factor $Y = X \exp(-\kappa\sigma/2)$, where

$$\kappa^2 = 4\pi l_B N_A \left( Z \frac{\rho_0}{M} \eta + 2c_s + 2c_b \right) \qquad (4)$$

is the square of the Debye screening length parameter $\kappa$, $N_A$ is Avogadro's constant, $\eta = \pi \rho \sigma^3 / 6$ is the protein volume fraction, $\rho$ is the number density of proteins of molar mass $M$, and $\rho_0$ is the protein mass density. The molar buffer and monovalent molar salt concentrations are denoted by $c_b$ and $c_s$, respectively. For the systems studied in this work, $\kappa$ is determined essentially by the added salt concentration. The geometric factor $X(\eta, \kappa\sigma)$ as obtained in the mean-spherical approximation (MSA) for pointlike microions is quoted in the Appendix. The factor $X$ accounts, within the linear MSA, for the reduced screening ability of the microions for nonzero concentration of proteins (macroions). It decreases with decreasing protein concentration and approaches the standard DLVO prefactor $1/(1+\kappa\sigma/2)$ for $\rho \to 0$. Since $Z$ is quite small, we have disregarded here the charge renormalization effect caused by quasicondensed counterions (see, e.g., Ref. 45). The reduced screening parameter $z_{rep} = \kappa\sigma$ quantifies the electrostatic screening length in units of $\sigma$. For later discussion, we abbreviate the nondimensionalized contact value of $u_{rep}(r)$ as $\beta\epsilon_{rep} = Z^2 l_B Y^2 / \sigma$.

In using this effective electrostatic interaction part, we neglect the discrete surface charge pattern of lysozyme. Model calculations by Allahyarov *et al.* in Refs. 33 and 46 indicate that multipolar pair interactions as well as correlations between the microionic co- and counterions due to their finite sizes are relevant for high salt concentrations only, i.e., typically when $c_s > 1.0$ mol/l. Under these high salt conditions, the electrostatic potential becomes radially nonmonotonic due to short-ranged, depletion-induced attractions.

Commonly, the $p$H-value and the excess amount of salt are carefully adjusted in a protein solution under experimental conditions. Then, the salt concentration $c_s$, the co- and counterion concentrations, and the protein net charges $Z$ are precisely known. Therefore, the electrostatic repulsive interaction part is completely determined by the system temperature $T$ and the protein volume fraction $\eta$.

To describe the attractive interactions between adjacent patches on two protein surfaces, we employ the patchy model description of Kern and Frenkel,[17] assuming that the radial and angular degrees of freedom can be factorized. The attractive interaction potential part $\tilde{u}_{attr}(r)$ hereby is angularly modulated by an angular distribution function $d(\Omega_1, \Omega_2)$ that depends on the solid angles $\Omega_1$ and $\Omega_2$ of two particles 1 and 2, respectively, according to

$$u_{attr}(r, \Omega_1, \Omega_2) = \tilde{u}_{attr}(r) \times d(\Omega_1, \Omega_2). \qquad (5)$$

The particles are assumed to have $\alpha = 1, \ldots, n$ attractive spherical caps on each surface, with an opening angle $\delta$ around the normal direction $\mathbf{e}_\alpha$ of each cap. Two particles, 1 and 2 (see Fig. 1), attract each other only if the center-to-center vector $\mathbf{r}$ intersects simultaneously the patchy areas on particle 1 and particle 2. This is equivalent to demanding for attraction that the angle $\theta_{12,\alpha}$ between a normal vector $\mathbf{e}_\alpha$ of patch $\alpha$ on particle 1, and the angle $\theta_{21,\beta}$ between a normal vector $\mathbf{e}_\beta$ of patch $\beta$ on particle 2 are simultaneously smaller

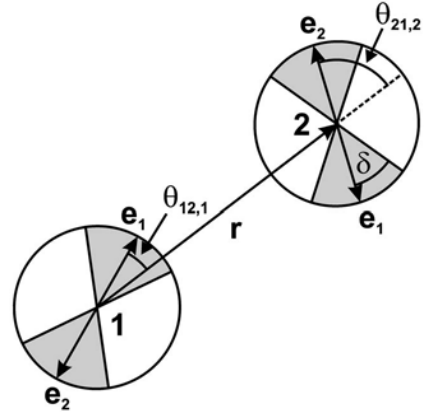

FIG. 1. Schematic drawing of a configuration of two model proteins, each carrying two attractive patches (gray areas). For the given configuration, the two particles repel each other according to the screened electrostatic potential in Eq. (3), with the surface charge assumed to be smeared out homogeneously over the sphere surface. There is no attractive interaction part since the center-to-center vector $\mathbf{r}$ does not intersect simultaneously the shaded attractive patches on particles 1 and 2. See the main text for the definitions of the remaining symbols.

than $\delta$. The angular distribution function $d(\Omega_1, \Omega_2)$ is thus given by

$$d(\Omega_1, \Omega_2) = \begin{cases} 1 & \text{if} \begin{cases} \theta_{12,\alpha} \leq \delta & \text{for a patch } \alpha \text{ on } 1 \\ \text{and } \theta_{21,\beta} \leq \delta & \text{for a patch } \beta \text{ on } 2 \end{cases} \\ 0, & \text{otherwise}. \end{cases}$$
$$(6)$$

Different from the work of Kern and Frenkel, where an attractive square-well potential has been used for $\tilde{u}_{attr}(r)$, we use here an attractive Yukawa-type potential of the form

$$\tilde{u}_{attr}(r) = \begin{cases} -\tilde{\epsilon}_{attr}(T) \sigma \dfrac{\exp[-z_{attr}(r/\sigma - 1)]}{r}, & r > \sigma \\ 0, & r \leq \sigma, \end{cases} \qquad (7)$$

where the temperature-dependent potential depth $\tilde{\epsilon}_{attr}(T)$ is described as[16]

$$\tilde{\epsilon}_{attr}(T) = \epsilon_{attr} \left( 1 + \psi \frac{[T_c - T]}{T_c} \right). \qquad (8)$$

Here, $T_c$ is the critical temperature of liquid-gas coexistence, and $\epsilon_{attr}$ and $\psi$ are two physical parameters, which will be determined from experimental data at the critical point (see later). Since the strength of the attractive potential part increases with decreasing $T$, the signs of $\psi$ and $\epsilon_{attr}$ are defined to be positive. For $\psi = 0$, the attractive potential part would be temperature independent. The temperature dependence in Eq. (8) constitutes a first-order Taylor-expansion around $T_c$. It can be considered as a simple approximation to the so far not well understood temperature dependence of the attractive hydrophobic interactions. The expansion around $T = T_c$ has been selected since the critical temperature is an experimentally well-assessed quantity.

The fraction $\chi$ of the sphere surface covered by the $n$ attractive patches is given by the surface coverage factor,[17]

$$\chi = n\,\sin^2\!\left(\frac{\delta}{2}\right). \tag{9}$$

Within the present patchy model, only the square of $\chi$ appears in the average of $u_{\text{attr}}$ over the angular degrees of freedom. The surface coverage factor $\chi$ is thus an additional independent parameter in our anisotropic patchy model, and our calculations do *not* depend on the actual local distribution of patches and their individual sizes. All details of the discrete character of the pair interactions are convoluted in the surface coverage factor $\chi$ due to this angular averaging. However, in place of $\chi$, one can use the opening angle $\delta$ as the adjustable parameter for $n$ fixed or, likewise, $n$ is taken as the adjustable and $\delta$ is fixed. Our calculations have been performed such that $\delta$ is the independent parameter for $n$ fixed to 2, as sketched in Fig. 1. With this choice, an isotropic attractive potential is recovered in the limit $\delta \to \pi$.

The second osmotic virial coefficient $B_2$ has the following form for an angular-dependent pair potential:

$$B_2(T) = -\frac{1}{2}\int d\mathbf{r}\,\langle \exp[-\beta u(r,\Omega_1,\Omega_2)] - 1\rangle_{\Omega_1,\Omega_2}, \tag{10}$$

where

$$\langle \cdots \rangle_{\Omega_1,\Omega_2} = \frac{1}{(4\pi)^2}\int\int \cdots d\Omega_1 d\Omega_2 \tag{11}$$

denotes an unbiased angular average. The reduced second virial coefficient $B_2^*$ is defined as the ratio of $B_2$ and the virial coefficient, $B_2^0 = 2\pi\sigma^3/3$, of hard spheres of diameter $\sigma$, i.e., $B_2^* = B_2/B_2^0$.

Lysozyme is approximately an ellipsoidal polypeptide with volume $v_0 = (\pi/6)\times 4.5\times 3.0\times 3.0$ nm$^3$.[10] In the present work, we treat the ellipsoidal-like polypeptide as a spherical particle of equal volume $v_0$, and effective diameter $\sigma = 3.4$ nm.[6,47,48]

## IV. HELMHOLTZ FREE ENERGY AND PHASE COEXISTENCE

In order to explore the phase diagram of lysozyme, we need to calculate the Helmholtz free energies of the fluid and solid phases. For this purpose, we employ the thermodynamic perturbation theory of Barker and Henderson[49] using hard spheres as the reference system. The Helmholtz free energy of the actual system is hereby expanded in powers of the interaction strength of the perturbational potential part, $u_p = u - u_0$, with the hard-sphere reference system indicated by the subscript 0,

$$f(T,\eta) = f_0(\eta) + f_1(T,\eta) + f_2(T,\eta) + \cdots. \tag{12}$$

We have nondimensionalized here the Helmholtz free energy, $F(N,V,T)$, of the proteins by the thermal energy $1/\beta = k_B T$ and the volume per particle, $v_0 = \pi\sigma^3/6$, according to $f = \beta F v_0 / V$, where $N$ is the number of particles in the system volume $V$. The first-order perturbation contribution to the free energy contains only pairwise interactions and is given by

$$f_1(T,\eta) = 12\eta^2 \frac{1}{\sigma^3}\int_\sigma^\infty dr\, r^2 g_0(r)\langle \beta u_p(r,\Omega_1,\Omega_2)\rangle_{\Omega_1,\Omega_2}, \tag{13}$$

where $g_0(r)$ is the radial distribution function of hard spheres of volume fraction $\eta$. In the solid phase, $g_0(r)$ is the orientationally averaged pair distribution function.

The second-order perturbation contribution $f_2$ contains three- and four-body distribution functions and includes fluctuations in the particle density. Unfortunately, these terms cannot be computed easily because of the complexity of these higher-order distribution functions. For this reason, we involve the macroscopic compressibility approximation developed by Barker and Henderson,[50] which involves only the pair distribution function and the isothermal compressibility $\chi_T$ of the reference system according to

$$f_2(T,\eta) = -6\eta^2 \left(\frac{\partial \eta}{\partial \Pi_0}\right)_T \frac{1}{\sigma^3}\int_\sigma^\infty dr\, r^2 g_0(r)$$
$$\times \langle [\beta u_p(r,\Omega_1,\Omega_2)]^2\rangle_{\Omega_1,\Omega_2}. \tag{14}$$

Here, $\chi_T/(\beta v_0) = 1/\eta(\partial\eta/\partial\Pi_0)_T$, where we have nondimensionalized the protein osmotic pressure $\tilde{\Pi}_0$ according to $\Pi_0 \equiv \beta\tilde{\Pi}_0 v_0$.

In the fluid phase, the reduced free energy of the hard-sphere reference system $f_0$ consists of the ideal gas part,

$$f_0^{\text{id}}(\eta) = \eta[\ln(\eta\Lambda^3/v_0) - 1], \tag{15}$$

with the thermal wavelength, $\Lambda = h/\sqrt{2\pi m k_B T}$, involving the protein mass $m$, Planck's constant $h$, and the interaction free energy part, which we describe by the Carnahan–Starling equation of state,[51]

$$f_0^{\text{CS}}(\eta) = \frac{4\eta^2 - 3\eta^3}{(1-\eta)^2}. \tag{16}$$

Solid lysozyme dispersions are known to have a tetragonal crystal structure.[52] Within our simplifying model, we have mapped the ellipsoidal-like shape onto a spherical one, which allows us to use for simplicity the hard-sphere reference system with a fcc solid phase. Existing schemes for $g_0$ in solids[53–56] have been developed and compared with Monte Carlo simulation data only for fcc and body-centered-cubic (bcc) lattices. For the excess Helmholtz free energy density of the fcc hard-sphere solid phase, we use Wood's equation of state,[57] namely,

$$f_0^{\text{solid}}(\eta) = 2.1306\eta + 3\eta\ln\!\left(\frac{\eta}{1-\eta/\eta_{\text{cp}}}\right) + \eta\ln\!\left(\frac{\Lambda^3}{v_0}\right), \tag{17}$$

where $\eta_{\text{cp}} = \pi\sqrt{2}/6$ is the fcc volume fraction for closed packing. The integration constant (i.e., the first term on the right-hand side) is obtained from the free energy density of a hard-sphere crystal calculated from Monte Carlo simulations at $\eta = 0.576$.[58] Note that different free energy expressions are used for the fluid and solid phases of the reference system since there is a symmetry change in going from one phase to the other. As the radial distribution function $g_0(r)$ in the liq-uid phase, we use the Verlet–Weis corrected[59,60] Percus–Yevick solution,[61,62] and the orientation-averaged pair distribution function of Kincaid[56] for the fcc crystal phase.

The second-order perturbation scheme outlined above has been widely used for various perturbation potentials and compared with simulation data. For example, it has been used for approximating the free energies of fluid or solid phases of particles with attractive[63] and repulsive[64] short-ranged pair potentials of Yukawa-type, and particles with polymer-induced depletion interactions.[65–67] As long as the contact value of the perturbation potential part is not much larger than $k_B T$, so that $u_p$ can be treated as a perturbation relative to the dominating hard sphere contribution, the perturbation scheme works decently well, provided $u_p$ is not too long ranged. In our calculations, the second-order perturbation term is typically 10 to 20 times smaller than the first-order contribution.

At fluid-solid phase coexistence, the two phases must be in thermal, mechanical, and chemical equilibrium. Since the coexisting phases are in thermal contact, the only two conditions determining the volume fractions of the coexisting fluid ($f$) and solid ($s$) phases are the equality of the osmotic pressure,

$$\Pi_f(T, \eta_f) = \Pi_s(T, \eta_s), \quad (18)$$

and chemical potentials,

$$\mu_f(T, \eta_f) = \mu_s(T, \eta_s), \quad (19)$$

with

$$\Pi(T, \eta) = \eta^2 \left( \frac{\partial (f(T, \eta)/\eta)}{\partial \eta} \right)_T$$

and

$$\beta \mu(T, \eta) = \left( \frac{\partial f(T, \eta)}{\partial \eta} \right)_T. \quad (20)$$

At sufficiently low temperatures, a liquidlike ($l$) and a gaslike ($g$) phases of high and low density, $\eta_l$ and $\eta_g$, respectively, coexist along the gas-liquid coexistence curve. The liquid-gas coexistence is metastable, however, with respect to a fluid-solid phase coexistence. Under gravity, the two fluid phases are separated by a meniscus, and particles and energy can pass through this interface. Equilibrium is achieved for equal osmotic pressures,

$$\Pi_g(T, \eta_g) = \Pi_l(T, \eta_l), \quad (21)$$

and chemical potentials,

$$\mu_g(T, \eta_g) = \mu_l(T, \eta_l), \quad (22)$$

of the coexisting phases.

The spinodal instability curve of diverging isothermal compressibility is determined by

$$\frac{\partial^2 f(T, \eta)}{\partial \eta^2} = 0. \quad (23)$$

The binodal and spinodal merge at the critical point (see later).

We have evaluated the improper integrals in the perturbation scheme using Chebyshev quadrature for the zonal part of $g_0(r)$ and Romberg quadratures for the remainder, where the perturbation pair potential has almost decayed to zero and the angular-averaged pair distribution is nearly constant. Higher-order derivatives of the free energy have been computed to machine precision accuracy using Ridder's implementation of Neville's algorithm. The phase coexistence curves have been determined using a Newton–Raphson method with line search (see Ref. 68 for the invoked algorithms).

## V. DETERMINATION OF THE ATTRACTIVE INTERACTION PARAMETERS

We proceed by first characterizing the yet unknown interaction parameters $z_{\text{attr}}$ in Eq. (7) and $\epsilon_{\text{attr}}$ in Eq. (8), and compute subsequently the equilibrium phase diagram of lysozyme for the experimentally scanned part of the $T$-$\eta$ plane. Aside from these two interaction parameters, there are two additional unknown parameters in our patchy sphere model, namely, the parameter $\psi$ in Eq. (8), which characterizes the temperature dependence of the depth of the attractive potential part, and $\delta$, the opening angle of the patches [see Eq. (9)], which determines the surface coverage factor $\chi$ for the given number $n = 2$ of patches.

In a first attempt to determine these parameters, one could try to fit the binodal, obtainable in principal from our model, to the experimental one. However, the complexity of the involved thermodynamic expressions renders this direct approach very tedious. For simplicity, we choose a simpler strategy and focus on a characteristic point in the phase diagram, namely, the critical point of the metastable gas-liquid protein phase coexistence. Right at the critical point, $u_{\text{attr}}(r)$ is determined by $z_{\text{attr}}$, $\epsilon_{\text{attr}}$, and $\delta$ alone since its depth becomes temperature independent [see Eq. (8)]. At the critical point, the second and third density derivatives of the Helmholtz free energy vanish, i.e.,

$$\frac{\partial^2 f(T_c, \eta_c)}{\partial^2 \eta} = 0$$

and

$$\frac{\partial^3 f(T_c, \eta_c)}{\partial^3 \eta} = 0. \quad (24)$$

Here, we use the critical temperature $T_c$ and the volume fraction at the critical point $\eta_c$ as determined experimentally. To obtain a third condition for the three unknown parameters, we exploit an empirical observation made by Warren[22] and Poon et al.[48] These authors find that the $B_2^*(T)$ of lysozyme is practically independent of the salt concentration for values larger than $c_s = 0.25$ mol/l, with a plateau value of $B_2^* = (-2.7 \pm 0.2)$. Hence, as an additional constraint, we demand that $B_2^*(T_c)$ is equal to

$$B_2^*(T_c) = -2.7. \quad (25)$$

This requirement is reasonable, since $B_2^*$ is the second term in the density expansion of the Helmholtz free energy density, $f(T, \eta) = f_0^{\text{id}}(\eta) + 4 B_2^*(T) \eta^2 + \mathcal{O}(\eta^3)$, so that any viable

TABLE I. System and pair potential parameters used in the thermodynamic perturbation calculation of the metastable gas-liquid binodal/spinodal, and the stable fluid-solid coexistence curve (for salt concentrations $c_s$ as indicated). The attractive potential part parameters $z_{attr}$ and $\beta\epsilon_{attr}$ are determined by Eqs. (24) and (25), respectively, using $Z=8$. For given $c_s$, the parameters $z_{attr}$, $\beta\epsilon_{attr}$, and $\delta$ (with a fixed value $n=2$) are determined from the experimental values for $\eta_c(c_s)$, $T_c(c_s)$, and $B_2^* = -2.7$, with $\psi$ fixed to 5.

| $c_s$ (mol/l) | $T_c$ (K) | $z_{rep}$ | $\epsilon_{rep}/(k_B T_c)$ | $z_{attr}$ | $\epsilon_{attr}/(k_B T_c)$ | $\delta$ (deg) | $\chi$ |
|---|---|---|---|---|---|---|---|
| 0.5 | 291.3 | 8.43 | 0.51 | 3.02 | 3.06 | 73.0 | 0.707 |
| 0.4 | 286.2 | 7.63 | 0.60 | 3.08 | 3.15 | 73.5 | 0.716 |
| 0.3 | 279.8 | 6.76 | 0.73 | 3.15 | 3.27 | 74.0 | 0.725 |
| 0.2 | 270.3 | 5.76 | 0.94 | 3.18 | 3.50 | 74.3 | 0.729 |

model should at least reproduce this value correctly.

Gibaud[69] have experimentally determined the reduced second virial coefficient of hen egg lysozyme as a function of $T$. These experimental data support Eq. (25) and show that there is a narrow band of $B_2^*(T_c)$ values, for which the fluid phase becomes unstable and separates into a gas- and liquid-like phase, in accordance with the extended principle of corresponding states suggested by Noro and Frenkel.[37] Foffi and Sciortino[70] have recently shown, using computer simulations, that the principle of corresponding states holds also for nonspherical symmetrically pair interaction potentials. Rosenbaum and Zukoski[9] have demonstrated that the solubility curves collapse onto a single master curve when plotted in the $B_2^*$-$\eta$ plane or, likewise, in the $\tau$-$\eta$ plane, where $\tau$ is the stickiness parameter in the adhesive hard-sphere model considered by them. Very recently, Gibaud has shown additionally on the basis of the present data set that the experimental gas-liquid curves of lysozyme suspensions superimpose for various salt concentrations when plotted in the $B_2^*$-$\eta$ plane. Such a scaling behavior of the gas-liquid coexistence curves is expected for systems interacting with short-ranged attractions because the term containing $B_2^*$ describes the mayor non-hard-sphere-like contribution to the Helmholtz free energy as we have noticed before. Therefore, $B_2^*$ can be only a crude measure of the actual form of the pair interaction potential and, as a consequence, is quite insensitive to small changes in the interaction parameters. A case in point will be the gas-liquid coexistence curves discussed in the following (see, especially Fig. 3).

The so far unknown parameters, $z_{attr}$, $\epsilon_{attr}$, and $\delta$, characterizing the attractive pair interaction part can now be obtained numerically from solving the set of nonlinear algebraic Eqs. (24) and (25). The additional free parameter $\psi$ in Eq. (8) mainly determines the width of the coexistence curve. Its value will be adjusted when we compare the calculated and experimental binodals and spinodals (see below).

At this point, we emphasize that the second-order perturbation term in Eq. (14) is a necessary contribution which allows to fix $\chi$ independently of $\epsilon_{attr}$. Carrying out the angular average results in a factor of $\chi^2$. When the first-order perturbation term is considered alone, $\chi^2$ and $\epsilon_{attr}$ appear only as a product. Thus, one cannot choose $\chi$ (or, respectively, $\delta$ at fixed $n$) and $\epsilon_{attr}$ independently when the first-order perturbation contribution to the free energy of the reference hard sphere system in Eq. (12) is considered only.

In the present second-order perturbation theory, density fluctuation effects are ignored, which in general lower the critical temperature. However, the fluctuations become less important with increasing range of the pair interactions[71] since the number of particles contributing to the force experienced by a central one increases with increasing range of attraction, so that the mean-field picture becomes more accurate (see, e.g., Fig. 1 in Ref. 72). Thus, we can expect that the fluctuation-induced shift of the critical point is rather small in lysozyme solutions, as argued earlier by Sear and Gelbart.[31]

The parameters determined by the evaluation strategy described above are summarized in Table I. Note that the range of the screened Coulomb repulsion $1/z_{rep}$ and its strength $\beta\epsilon_{rep}$ show the expected increase with decreasing salt concentration. The temperature dependency of the Bjerrum length, through $\varepsilon(T)$, has been accounted for. However, in the considered temperature range, $l_B$ is only mildly dependent on $T$.

Due to the stronger electrostatic repulsion between the proteins on lowering the salt concentration, $T_c$ decreases with decreasing salt concentration. Figure 2 shows the repulsive potential part $u_{rep}(r)$, the angular-averaged attractive potential part $\langle u_{attr}(r) \rangle_{\Omega_1,\Omega_2}$, and the angular-averaged total perturbation potential $\langle u_p(r,\Omega_1,\Omega_2) \rangle_{\Omega_1,\Omega_2}$, obtained at the critical concentration for $c_s = 0.5$ mol/l. Note that the contact value, $\epsilon_{\text{attr}}/(k_B T_c)$, of the *non-angular-averaged* attractive pair potential at $T_c$ given in Table I, is well above $3k_B T_c$. In contrast, Figs. 2 and 3 show the *angular-averaged* attractive interaction part, with contact value $\langle u_{\text{attr}}(\sigma)/(k_B T_c) \rangle_{\Omega_1,\Omega_2} = \chi^2 \epsilon_{\text{attr}}/(k_B T_c)$. Thus, the angular-averaged contact value of the attractive part is smaller than $3k_B T_c$ (see Figs. 2 and 3).

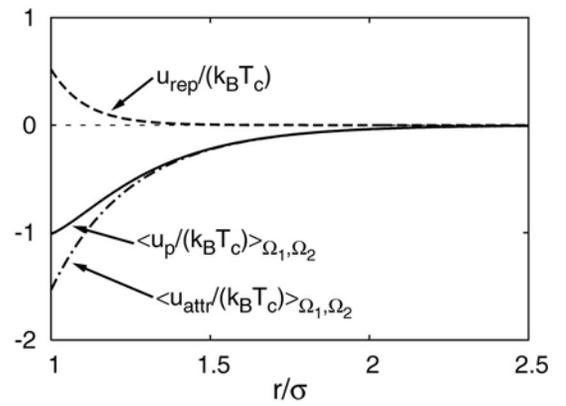

FIG. 2. Repulsive electrostatic pair potential part $u_{rep}/(k_B T_c)$ (thick dashed curve), angular-averaged attractive interaction part $\langle u_{attr}/(k_B T_c) \rangle_{\Omega_1,\Omega_2}$ (thick dashed-dotted curve), and total perturbational pair potential $\langle u_p/(k_B T_c) \rangle_{\Omega_1,\Omega_2}$ (thick solid curve) for parameters at the critical point, where $c_s = 0.5$ mol/l, using $\eta_c = 0.17$, $T_c = 291.3$ K. The parameters used in the perturbational interactions for the attractive and repulsive Yukawa-type potential parts are listed in Table I. At larger $r$, $\langle u_p/(k_B T_c) \rangle_{\Omega_1,\Omega_2}$ is dominated by the attractive interaction part.

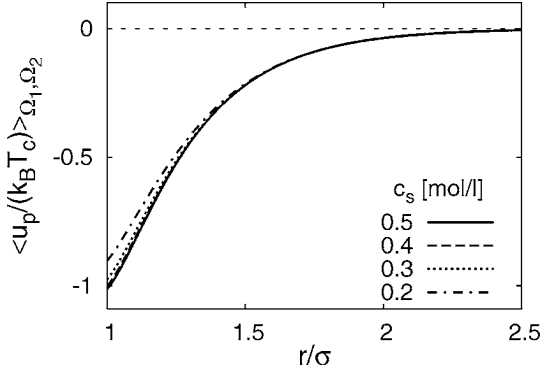

FIG. 3. Angular-averaged total perturbation potential, $u_p = u - u_0$ [see Eq. (1)] for various salt concentrations as indicated. With decreasing $c_s$, the contact value of $\langle u_p/(k_B T_c) \rangle_{\Omega_1,\Omega_2}$ decreases due to the enlarged range of the electrostatic repulsion part.

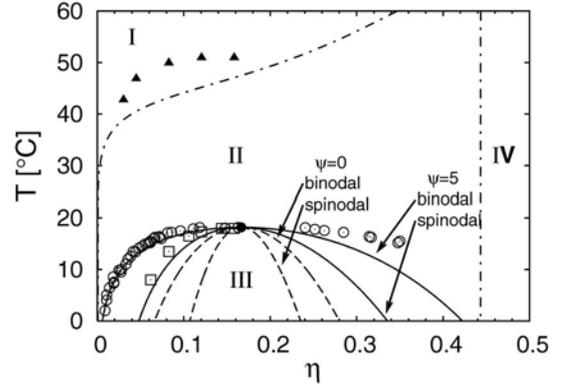

FIG. 4. The phase diagram of aqueous lysozyme solutions for $c_b = 0.02$ mol/l HEPES buffer, and pH=7.8, $c_s = 0.5$ mol/l NaCl. The circles (○) describe the experimentally found metastable gas-liquid coexistence curve (Ref. 38), the squares (□) indicate the spinodal (Ref. 38), and the black triangles (▲) depict the experimental fluid-crystal coexistence curve. The two dashed curves show the calculated binodal and spinodal, respectively, for $\psi = 0$. The two solid curves describe the calculated binodal and spinodal, respectively, where the two curves account for an additional temperature dependence of the attractive potential depth with $\psi = 5$ [see Eq. (8)]. The dashed-dotted curves are the calculated fluid-crystal coexistence curves for $\psi = 5$, with the interaction parameters determined from the experimental data at the critical point as explained in the text. In region I, a stable fluid phase is observed, whereas one finds a fluid-crystal coexistence in region II, a metastable gas-liquid coexistence in region III, and a pure crystalline phase in region IV.

The range $1/z_{\text{attr}}$ and depth $\beta\epsilon_{\text{attr}}$ of the attractive Yukawa potential exceed the range $1/z_{\text{rep}}$ and strength $\beta\epsilon_{\text{rep}}$, respectively, of the repulsive part, so that the averaged perturbation pair potential is purely attractive. Actually, this finding holds true for all salt concentrations considered, as can be noticed from Fig. 3. Due to the weaker screening of the protein surface charge at lower salt content $c_s$, the attraction range of the total potential decreases with decreasing amount of salt. However, in contrast to the drastic change of the repulsive interaction part with $c_s$, the parameters of the attractive potential part vary only slightly with the salinity. The range of attraction $1/z_{\text{attr}}$ shrinks by 6% only when $c_s$ is reduced from 0.5 to 0.2 mol/l, whereas the attraction strength $\beta\epsilon_{\text{attr}}$ increases by 14%. According to our calculations, the opening angle $\delta$, and thus the surface coverage $\chi$, increase only slightly with decreasing salinity. These changes in $\chi$ and $\delta$ are negligible as compared to the strong influence of the salinity on the electrostatic screening length. Therefore, we can conclude that in our model the range and strength of the radially averaged attractive potential part are approximately constant within the salt range considered.

We have carefully checked the sensitivity of the calculations to small changes in the employed parameters. Changing $B_2^*$ from $-2.7$ to $-2.5$ or, likewise, to $-2.9$, and keeping all other parameters unchanged, leads to changes in $z_{\text{attr}}$ and $\beta\epsilon_{\text{attr}}$ by less than 5%, whereas the surface coverage factor is affected by 2% only. Varying the bare protein charge number $Z=8$ by $\pm 2$, keeping again all other parameters fixed, changes both $z_{\text{attr}}$ and $\beta\epsilon_{\text{attr}}$ by less than 6%, and $\chi$ by less than 3%. As expected, our calculations are more sensitive to variations in the critical volume fraction: Assuming an uncertainty of 10% in the experimental $\eta_c$, say $\eta_c = (0.17 \pm 0.02)$, $z_{\text{attr}}$ changes by up to 29%, whereas $\beta\epsilon_{\text{attr}}$ is changed by 6%, and $\chi$ by 3%. An uncertainty in the protein diameter of $\pm 0.2$ nm (Ref. 7) causes deviations in $z_{\text{attr}}$ and $\beta\epsilon_{\text{attr}}$ by less than 4%, and in $\chi$ by less than 2%.

## VI. CALCULATED PHASE DIAGRAMS

In Fig. 4, the phase diagram is shown for the largest salt concentration considered of $c_s = 0.5$ mol/l. As can be seen from this figure, the predicted gas-liquid coexistence curves are too narrow when $\psi = 0$ is used (dashed curves). To correct for this, we have introduced the temperature-dependent coupling parameter $\tilde{\epsilon}_{\text{attr}}$ in Eq. (8), which includes the parameter $\psi$. Positive values of $\psi$ widen the unstable region in the calculated phase diagram because of the increase in the strength of attraction. From calculating the binodals (solid curves) for a variety of $\psi$ values, and comparing them with the experimental data points at $c_s = 0.5$ mol/l, we find good agreement, using a value $\psi = 5$, for all volume fractions smaller than 20%. At larger volume fractions, the calculated binodals deviate somewhat from the experimental ones. We note, however, that changing $\psi$ by not more than 40% does not crucially effect the overall good agreement between experimental and calculated binodals and spinodals.

The range $1/z_{\text{attr}}$ and the strength $\beta\epsilon_{\text{attr}}$ of the attractive potential part, obtained for one specific salt concentration ($c_s = 0.5$ mol/l) at the critical point has been fixed in calculating the coexistence curves also for the other values of $c_s$ considered. The binodal and the fluid-crystal coexistence curves have been calculated according to the double tangent construction using Eqs. (18) and (19). The spinodal curve follows from the condition that the isothermal compressibility diverges [see Eq. (23)].

In Fig. 5 finally, the calculated gas-liquid coexistence curves are shown for four different salt concentrations in comparison with the experimental data points. We could have adjusted the parameter $\psi$ for each $c_s$ separately. However, we find that fixing it to $\psi = 5$ results in binodals that describe the experimental ones quite well for all salinities. For each $c_s$ considered, the binodal curve is described reasonably well for low volume fractions, whereas as discussed before, our model underestimates the transition temperature systematically at higher protein concentrations. This might be due to salt partitioning over the two phases which is not accounted for in our model calculations.[7,22]

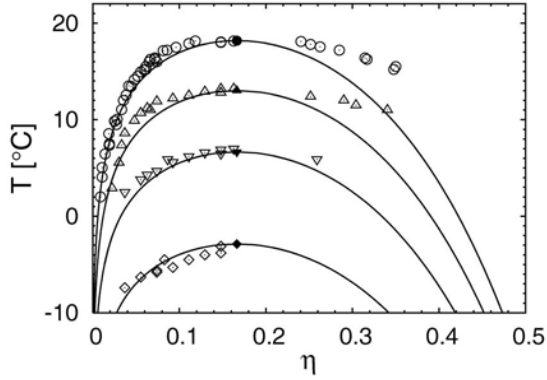

FIG. 5. Gas-liquid coexistence curves of a lysozyme solution obtained experimentally from temperature quenches at four different salt concentrations: $c_s$=0.5 mol/l (○), 0.4 mol/l (△), 0.3 mol/l (▽), and 0.2 mol/l (◇). The filled symbols mark the critical points estimated from the experiment. The solid curves describe the coexistence curves as calculated from our model using a fixed value $\psi$=5.

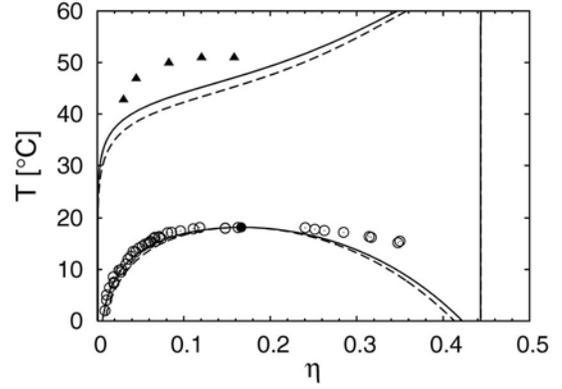

FIG. 6. Phase diagram of lysozyme for $c_s$=0.5 mol/l NaCl, $c_b$=0.02 mol/l HEPES buffer, and $p$H=7.8. The symbols indicate the experimental data points identical to the ones in Fig. 4. The solid curves describe the equilibrium phase diagram obtained from the anisotropic model. For comparison, the dashed lines describe the gas-liquid and fluid-solid coexistence curves are obtained from a purely isotropic pair potential. In both cases, $\psi$ is set equal to 5.

## VII. DISCUSSION

The virtue of our model potential as compared to using a square-well potential alone[15,16,18] is that we account explicitly for the screened electrostatic repulsion. Through this model extension, we can distinguish the influence of the excess salt concentration from the attractive potential part that is not well-understood in its details. For the attractive part, in turn, we have adopted a simplifying patchy model of Yukawa-type in its radial factor. We have determined the interaction parameters of $u_{attr}$ from the experimental data for the $T_c$, $\eta_c$, and $B_2^*(T_c)$ of lysozyme at the critical point. Using the experimental data at the critical point, we find a range of attraction of $0.33\sigma$ at $c_s$=0.5 mol/l, and $0.31\sigma$ at $c_s$=0.2 mol/l. These ranges of attraction are consistent with the corresponding findings of several authors as summarized by Lomakin et al. (see Fig. 5 in Ref. 16). In fact, such an intermediate range of attraction is very remarkable since in the case of a *solely* isotropic attractive potential of Yukawa-type, one expects a stable gas-liquid coexistence region for ranges of attraction above $0.17\sigma$ (or, correspondingly, for $z_{attr} < 6$).[63,73] In the present case of an attractive and repulsive pair interaction potential of Yukawa-type, the fluid phase is stabilized against gas-liquid phase separation due to the charge-induced electrostatic repulsion, which shifts the gas-liquid critical point below the solubility curve, and thus, leads to a metastable binodal. We note that the range of attraction of $(1.0 \pm 0.1)$ nm $(\sim 0.3\sigma)$ experimentally found by Israelachvili and Pashley[74] measuring the force between two hydrophobic plates is in excellent accord with our findings.

In a number of previous studies, the isotropic and rather short-ranged DLVO pair potential has been used to fit the experimental scattering data on lysozyme.[4,6,47] To make contact with this earlier work, consider now a purely isotropic pair interaction by setting $\chi$=1 in our model. In the isotropic case, we obtain $1/z_{attr}$=0.36, using the same method to determine the attractive part as in the nonisotropic case. This attraction range, in fact, is nearly identical to the one observed for the anisotropic case since the critical volume fraction depends only weakly on the patchiness.[17] On the other hand, the potential depth for $\chi$=1 is given by $\beta\epsilon_{attr}$=1.39, which corresponds to $B_2^*(T_c)$=−1.26. This value for $B_2^*(T_c)$, obtained from assuming isotropic attractions, disagrees strongly with the experimentally observed value $B_2^*(T_c)$=−2.7. In contrast, our patchy model for $\chi < 1$ is capable of describing the experimental data, and it accounts for the influence of the added salt.

In Fig. 6, we compare the gas-liquid and fluid-solid coexistence curves, for an isotropic interaction potential with $\chi$=1, with the results from our anisotropic model from Fig. 4. As can be seen, the fluid-solid coexistence curve is shifted only slightly to lower temperatures when an isotropic pair interaction potential is assumed. Hence, isotropic attractive pair interactions for the protein solution result in a solubility curve located further below the experimental data, for interaction parameters determined again at the experimental critical point. Even in the isotropic case, the gas-liquid coexistence curve remains metastable relative to the fluid-solid coexistence curve also in the isotropic case, which might be due to the effect of the competing repulsive and attractive interactions. Such a weak influence of the patchiness on the location of the coexistence curves is expected in our model calculations since only the orientationally averaged pair potential enters into the free energy expression. In fact, the angular-averaged contact value $\langle\beta\epsilon_{attr}\rangle_{\Omega_1,\Omega_2}=\chi^2\beta\epsilon_{attr}$=1.53 (see Table I), observed using an anisotropic pair interaction potential, differs only slightly from the contact value, $\beta\epsilon_{attr}$=1.39, for the isotropic pair potential. However, the fact that the calculated $B_2^*$ for an isotropic interaction potential disagrees by a factor of 2 with the experimentally observed virial coefficient and the observation that the fluid-solid curve is located further below the experimental data than the one for anisotropic interactions implies that the experimental data can be consistently described only for an anisotropic pair interaction. Furthermore, our phase boundary calculations for isotropic versus anisotropic interactions highlights why in earlier calculations on the phase behavior of lysozyme, based on assuming a short-ranged isotropic pair potential, qualitative agreement with the experimental data has been achieved. In fact, aside from the totally wrong prediction for $B_2^*(T_c)$, an isotropic attractive pair potential can result in a reasonably good qualitative agreement with the experimental phase coexistence curves.

Carpineti et al.[35] discuss the need to account for hydrophobic patches in order to explain the temperature dependence of the solubility curve. Our model calculations conform their suggestion since the experimental data are recovered with the correct $B_2^*$ only for $\chi \neq 1$. Curtis et al.[36] have argued that 51% of the lysozyme surface area is hydrophobic, a value not too different from the surface coverage factor found in our work (we obtained $\chi = 71\%-73\%$). In addition, Curtis et al.[36] concluded from their experimental data that the nonpolar (hydrophobic) area on the protein surface decreases by the addition of sodium chloride (see Table I), in agreement with our findings. Our phenomenological description of the hydrophobic interactions between adjoined patches using a Yukawa-type attractive interaction potential part indicates that these interactions are only slightly affected by the salt concentration. All the experimental binodals for $c_s = 0.2, 0.3, 0.4$, and $0.5$ mol/l can be well described using a fixed value of $\psi = 5$ (see Fig. 5). Only the prefactor, $\epsilon_{attr}$, and $\chi$ decrease slightly with increasing $c_s$ (see Table I). Thus, the main effect of salt is to screen the lysozyme net charges as expected.

To arrive at a physical understanding of the strong temperature dependence of the attractive interaction part, as indicated in lysozyme solutions by a nonzero value of $\psi = 5 \pm 2$ (see Sec. VI), is a demanding task since little is known about the underlying molecular mechanism.[75]

Some progress on the microscopic understanding of the attractive interactions has been made only very recently by Horinek et al.[76] Their main observation is that the force between two hydrophobic objects is caused by two contributions of comparable strength; namely, van der Waals attractions and water-structure effects. Because the van der Waals attractions are to a first approximation temperature independent on neglecting the trivial temperature dependence due to the Boltzmann weight of the Hamiltonian, we attribute the strong temperature dependence in lysozyme solutions, indicated by a nonzero value of $\psi = 5 \pm 2$, mainly to the change in the water structure close to the hydrophobic surface.[77] Lomakin et al.,[16] who used an aeolotopic model to describe the phase behavior of $\gamma$-crystalline protein solutions, have arrived earlier at a similar conclusion regarding the strong temperature dependence of the attractive interactions (see p. 1655 in Ref. 16). Furthermore, they found a comparable value of $\psi = 3$ in $\gamma$-crystalline protein dispersions. These authors propose alternatively that the extended width of the gas-liquid coexistence might also be due to the discrete and anisotropic character of the hydrophobic interactions. Using computer simulations, Kern and Frenkel[17] showed that the gas-liquid coexistence curves can broaden significantly for sufficiently short-ranged attractive pair potentials and low surface coverage. Thus, we cannot see within our simple model as to whether the broadening of the gas-liquid coexistence curve is due to a strong temperature dependence or due to the patchiness.

Understanding protein crystallization is a complex issue. The dashed-dotted fluid-crystal coexistence curve in Fig. 2 deviates to some extent from the experimental data at higher volume fractions. However, aside from this, the calculated phase diagram agrees qualitatively with the experimental one regarding the metastability of the gas-liquid coexistence curve, and the extent of the gap between the critical point and the fluid-solid coexistence curve. In recent work,[15,78,79] it has been demonstrated that the specific geometry, i.e., the number of patches, their size and their distribution across the surface, significantly affects the ability to form crystals, the nucleation kinetics and the crystalline order. In particular, crystallization is expected to be hindered whenever the preferred local order in the liquid state is incompatible to the crystalline space symmetry. One speaks then of a "frustrated" liquid state.[79] In this case, the pair potential is no longer angularly averageable to describe the solid state.[15] Furthermore, McManus et al.[80] have shown for human $\gamma$D-crystalline proteins, that angular averaging is a feasible simplification to describe the fluid phase in its dependence on the number of spots on the protein surface, whereas the discrete patchiness influences crucially the solubility curve. One can speculate that this applies also for lysozyme dispersions.

**VIII. CONCLUSION**

We have studied the phase behavior of lysozyme dispersions on the basis of a pair potential consisting of a repulsive DLVO-type screened Coulomb part plus a patchy attractive part.

The strength and the range of the attractive radial potential factor of Yukawa-type, and the surface coverage of patches, have been determined using the experimentally known values for the concentration, temperature, and reduced second virial coefficient of lysozyme at the gas-liquid critical point. With the so determined patchy pair potential, we have calculated the metastable gas-liquid coexistence and spinodal curves of lysozyme solutions, and the fluid-solid coexistence curve, using the compressibility approximation of second-order thermodynamic perturbation theory. The shape of the computed phase diagram conforms overall quite well with the experimental data, in particular regarding the salt dependence of the coexistence curve, and the width of the gap in between the binodal and the fluid-solid coexistence curves. The percentage of surface coverage of patches ($\sim 70\%$) obtained in our model, and the interaction range of about 30% of the diameter, and the temperature dependence of the attractive interaction part, as well as the salt dependence of the interaction strength, are consistent with previous findings. This consistency is encouraging and supports the applicability of our simple model to describe lysozyme solu-tions. To obtain the solubility curve more accurately, however, might require to account for the patch geometry explicitly, without invoking an orientational preaveraging.

## ACKNOWLEDGMENTS

This project has been partly supported by the European Commission under the 6th Framework Program through integrating and strengthening the European Research Area. Contract No. SoftComp, NoE / NMP3-CT-2004-502235.

## APPENDIX: ONE-COMPONENT MACROION-FLUID POTENTIAL

Belloni[44] provides an analytical expression for the effective pair potential in Eq. (3) using the MSA and assuming pointlike microions. Within this approximation, the DLVO potential part is corrected by a factor $X$ depending on the reduced inverse screening length $\kappa\sigma/2$ and the macroion volume fraction $\eta$ according to

$$X = \cosh(\kappa\sigma/2) + U[\kappa\sigma/2 \cosh(\kappa\sigma/2) - \sinh(\kappa\sigma/2)], \quad (A1)$$

where

$$U = \frac{z}{(\kappa\sigma/2)^3} - \frac{\gamma}{\kappa\sigma/2} \quad (A2)$$

and

$$\gamma = \frac{\Gamma\sigma/2 + z}{1 + \Gamma\sigma/2 + z}, \quad (A3)$$

with $z = 3\eta/(1-\eta)$. The MSA screening parameter $\Gamma$ is uniquely obtained from solving the following relation:

$$\Gamma^2 = \kappa^2 + \frac{q_0^2}{(1 + \Gamma\sigma/2 + z)^2}, \quad (A4)$$

where $q_0 = \sqrt{4\pi l_B \rho Z}$. In the infinite dilute limit, $\rho \to 0$, $\Gamma$ reduces to the inverse Debye screening length $\kappa$. For an extension of Belloni's expression to differently sized and charged colloidal spheres, see Ref. 81.